# Thermal antenna behavior for thin films structures

Philippe Ben-Abdallah

Centre National de la Recherche Scientifique, Ecole Polytechnique, Laboratoire de Thermocinétique,

44306 Nantes, France

We show that under proper conditions thin films structures behave like thermal sources that are able to emit radiation in well defined directions over a broad spectral band for both polarization states of light. This effect results from the quantization of modes inside the structure as in a Fabry-Perot resonator. A theoretical demonstration of this effect is given using the matrix transfer method. This result is similar to the best efficient reported results with gratings but using a completely different physical principle [J. J. Greffet et al., Nature **416**, 61 (2002)].

A thermal source is often presented as a quasi-isotropic object. Recently, it has been recognized[1] for polar materials surmounted by a periodic grating that an emission in narrow angular lobes could be observed. This effect which is the signature of a high degree of spatial coherence for the thermal light has been shown to be the result of the surface phonon-polariton diffraction by the grating. However such a behavior is based on physical principles which strongly limit its applications in the field of thermal transfers. Indeed, for 1D grating, no lobe of emission can be observed in s-polarization since, in this case, surface waves cannot exit. Then, energy radiated by the source is localized in well defined directions only for p-polarization, the emissivity remaining globally isotropic for s-polarization. Another direction has been recently explored to design antenna from left-handed material[2] constructed with periodic metallic structures. Near plasmon resonance, the effective optical index of LHM is close to zero. Therefore, in accordance with the Snell-Descartes laws, the radiation emitted by a source embedded inside this medium is refracted around its normal. However, once again this effect is limited to TM waves. Although these results make possible to consider many applications at localized frequencies they seem much more difficult to exploit for designing spatially coherent thermal sources over a broad spectral band.

In the present paper we demonstrate that, under proper conditions, quantum size effects in absorbing thin films can give rise to an antenna behavior for both polarization states and over a relatively large spectral band. It has been recognized for some time with the semiclassical Lamb theory of laser[3-4] that absorbing media are able to support the presence of metastable waves. In bounded media such waves are quantified and can be stimulated by external propagative waves. Their interaction are described by the theory of the partially coherent light[5-6] when the film thickness is of the same order than the coherence length of light. In this paper we suggest that thin films can be used to designed thermal sources that are able to emit in narrow solid angles as a antenna would do it. To our knowledge, so far, this effect has never been analyzed.

As starting point one considers a homogeneous film of absorbing dielectric of thickness d (fig. 1) surrounded by a transparent dielectric. The coordinate system is chosen so that the yz plane is parallel to the surfaces of the film. A convenient framework for studying such a system is provided by the networks theory. It consists in calculating the transfer matrix (or diffusion matrix) of system. The magnitude of incident and reflected waves on both sides of film are related by

$$\begin{pmatrix} E^{in} \\ E^{out} \end{pmatrix}_{-} = T_{12} T_{2}^{prop} T_{21} \begin{pmatrix} E^{out} \\ E^{in} \end{pmatrix}_{+}, \qquad (1)$$

where $T_{ij}$ and $T_{i}^{prop}$ stand by the transfer matrix across the interface i-j and the propagation matrix through medium i, respectively. They are basically defined in terms of Fresnel transmission and reflection coefficients as follows

$$T_{ij} = \frac{1}{t_{ij}} \begin{pmatrix} 1 & r_{ij} \\ r_{ij} & 1 \end{pmatrix}, \qquad (2)$$

$$T_{j}^{prop} = \begin{pmatrix} e^{if_j} & 0 \\ 0 & e^{-if_j} \end{pmatrix}, \qquad (3)$$

where $f_j$ represents the phase parameter along the film. Without external excitation no incident wave highlights the film so that

$$E_{-}^{in} = E_{+}^{in} = 0. \qquad (4)$$

In these conditions the dispersion relation of resonant modes of system is simply given by

$$\det \begin{pmatrix} 1 & -r_{21} & 0 & 0 \\ & (I) & & (T_2) \\ 0 & 0 & r_{21} & -1 \end{pmatrix} = 0. \qquad (5\text{-a})$$

This expression simplifies into the short form

$$1 - r_{12}^2 e^{-2ik_{2x}d} = 0. \qquad (5\text{-b})$$

Here we consider the waves defined by

$$K_{2x} = \frac{pm}{d}, \qquad (6\text{-a})$$

$$K_{1x} = 0, \qquad (6\text{-b})$$

as solutions of Eqs. (5) [m is a positive integer]. These modes correspond to the laser oscillations of the system. Here, it is important to outline that these solutions are independent on the polarization state. Since the film is dissipative, resonant modes are metastable that is they have a finite lifetime.

This time is given by the imaginary part of their complex pulsation $\bar{w}_m$. From their definition, the longitudinal component (6) of wave vector can equivalently be written

$$K_{2x}^2 = e_2 \frac{\bar{w}_m^2}{c^2} - K_{//}^2, \qquad (7\text{-}a)$$

$$K_{1x}^2 = e_1 \frac{\bar{w}_m^2}{c^2} - K_{//}^2. \qquad (7\text{-}b)$$

Comparing Eqs. (6) and (7) it immediately follows that the transverse component of wave vector (which is conservative in accordance with the Snell'law) writes

$$K_{//} = \sqrt{e_1}\, \frac{\bar{w}_m}{c} \qquad (8)$$

while the complex pulsation of $m^{th}$ mode expresses in terms of dielectric constants

$$\bar{w}_m = \frac{c p}{d} \frac{m}{\sqrt{e_2 - e_1}}. \qquad (9)$$

When an external wave excites the system from one of its sides the transmitivity and reflectivity are given from the Airy's formula

$$t \equiv \left(\frac{E_+^{out}}{E_-^{in}}\right)_{E_+^{in}=0} = \frac{t_{12} t_{21} e^{-ik_{2x}d}}{1 - r_{12}^2 e^{-2ik_{2x}d}}, \qquad (10\text{-}a)$$

$$r \equiv \left(\frac{E_-^{out}}{E_-^{in}}\right)_{E_+^{in}=0} = \frac{r_{12}(e^{-2ik_{2x}d} - 1)}{1 - r_{12}^2 e^{-2ik_{2x}d}}. \qquad (10\text{-}b)$$

Obviously the poles of expressions (10) are not located on the real frequency axis since the reflection coefficient $R = |r|^2$ and transmission coefficient $T = |t|^2$ always remain bounded. Nevertheless R and T values are directly influenced (as highlighted on fig. 2 for a single $SiO_2$ layer[7] 10 thick embedded into a germanium matrix) by its complex poles.

From the Kirchoff law the 'real' emissivity is obtained from the energy conservation law

$$e(w, k_{//}) = 1 - T - R \qquad (11)$$

with $w$ and $k_{//}$, the real pulsation and the real transverse component of emerging wave vector, respectively. Introducing the complex angles made by the wave on both side of the film with respect to the normal we have

$$k_{//} = \frac{w}{c}\sqrt{e_1}\sin q_{1m} = \sqrt{\left(\frac{w}{c}\right)^2 e_2 - k_{2x}^2}. \qquad (12)$$

If we identify $k_{2x}$ to $K_{2x}$ then from Eqs. (6) we have also

$$k_{//}^2 = \frac{w}{c}e_2 - (\frac{pm}{d})^2. \qquad (13)$$

Thus the emerging angle of radiation produced by the resonant modes is given by the following expression

$$\sin^2 q_{1m} = \frac{1}{e_1}[e_2 - (\frac{pmc}{dw})^2]. \qquad (14)$$

A comparison between the location of peaks on the fringes pattern of the emissivity curve performed from a direct calculation with the Kirchhoff's law (Eq. (11)) and the angle defined by Eq. (14) clearly reveals that external radiation fields are able to excite the leaky waves and makes them quasi-resonant. Moreover as confirmed by the simulations, these modes seem much better excited when the imaginary part of angles (14) is small. In fact this quantity is an increasing function of modes. This explains why the lobe of emission corresponding to the fundamental mode (m=1) is narrowest. A direct inspection of relation (14) shows also that the number N of discrete modes is finite inside a film and is given by

$$N = E\{\frac{2nd}{l}\}, \qquad (15)$$

where the symbol E stands by the integer part of the expression inside the bracket. In the light of the above arguments, we now show this resonance property can be used to built an infrared antenna over a broad spectral band without making use of nanostructured grating. According to expression (14) the condition for spectral invariance of $m^{th}$ mode is given by

$$\frac{dq_{1m}}{dw} = 0. \qquad (16)$$

A straightforward calculation leads to

$$e_2(w) = Ae_1(w) + (\frac{pcm}{d})^2 \frac{1}{w^2} \tag{17}$$

with

$$A = \frac{1}{e_1(w_0)}[e_2(w_0) - (\frac{pcm}{d})^2 \frac{1}{w_0^2}]$$

$w_0$ being the lowest pulsation. When the surrounding medium is gray on the spectral band $(w_0, w)$ then the dielectric constant of film must satisfy

$$e_2(w) = e_2(w_0) + (\frac{pcm}{d})^2 [\frac{1}{w^2} - \frac{1}{w_0^2}]. \tag{18}$$

Notice that condition (16) cannot be fulfilled for two modes at the same time. To fix the number of emission peak we have to adjust the film thickness. For a single peak, relation (15) shows this thickness must be chosen in the interval

$$\frac{l_0}{2n(l_0)} \leq d < \frac{l_0}{n(l_0)} \quad . \tag{19}$$

The direction of emissivity peak can be deduced from relations (14) and (19)

$$q_1 = Arc\sin[\frac{1}{n_1}\sqrt{n^2(l_0) - \frac{l_0^2}{4d^2}}]. \tag{20}$$

An example of thermal antenna for a phantom material satisfying Eq. (18) has been plotted on Fig. 3. The peak of emission at $q_1 = 12°$ has an angular width at half height as narrow as $2°$ and does not depend on the observation frequency. In these conditions, the angle of emission $q_{ext} \equiv \arcsin(n_1 \sin q_1)$ outside of the germanium matrix, assumed immersed into vacuum, is about $56°$. We have shown that numerous material behave similarly at different frequencies along the spectrum. To mention some of them, let us quote for example ZnS films[7] at cryogenic temperature or the glass films in the middle infrared.

In conclusion, we have presented an alternate way to the textured materials of producing highly directional thermal sources from thin dielectric films. Contrary to material surmounted by a surface grating a thin absorbing film is able to behave like an antenna simultaneously for both states of polarization. This behavior requires films with $\text{Re}[e(w)] = O(w^{-2})$ and $\text{Im}[e(w)] = O(1)$. We can expect than simple thin film devices open new prospects in the radiative properties design. This is an area which needs to be explored further.

Benabdallah@let.ensma.fr

# Caption

Fig. 1. Geometrical configuration of system.

Fig. 2. Spectral emissivity of a $SiO_2$ film of thickness $d = 10\mu m$ embedded into a germanium matrix for s-polarization. Comparison between the directions of emissivity peaks (the thin fringes on the spectrum) calculated from the Kirchhoff's law and the direction of quantified metastable states obtained with Eq. (20) shows that propagative waves coming from the surrounding make resonant the quantified modes within the film. The spectral emissivity for p-polarization (not shown here) is similar to this spectrum. Inset: emissivity curve versus the angle (in degree) for s-polarization at $\lambda = 6.6\ \mu m$. The peaks are due to fundamental and secondary modes, respectively.

Fig. 3. Spectral emission of a thin film (thickness: $7\mu m$ ; surrounding: germanium) designed to radiate as an infrared antenna. The optical properties of medium are given by expression (18) with $\varepsilon_2(\omega_0) = 0.8 + 0.06i$ (the p-polarization contribution has not been plotted). Within the inset are plotted the emissivity curves versus the angle (in degree) for s-polarization state at $7\mu m$ (solid line), $8\mu m$ (dotted line) and $9\mu m$ (dashed line), respectively.

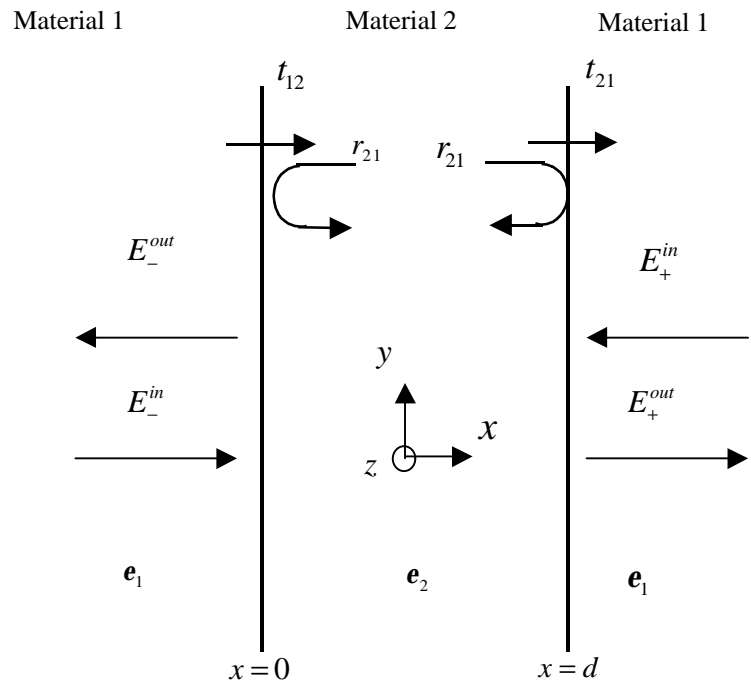

**Figure 1**

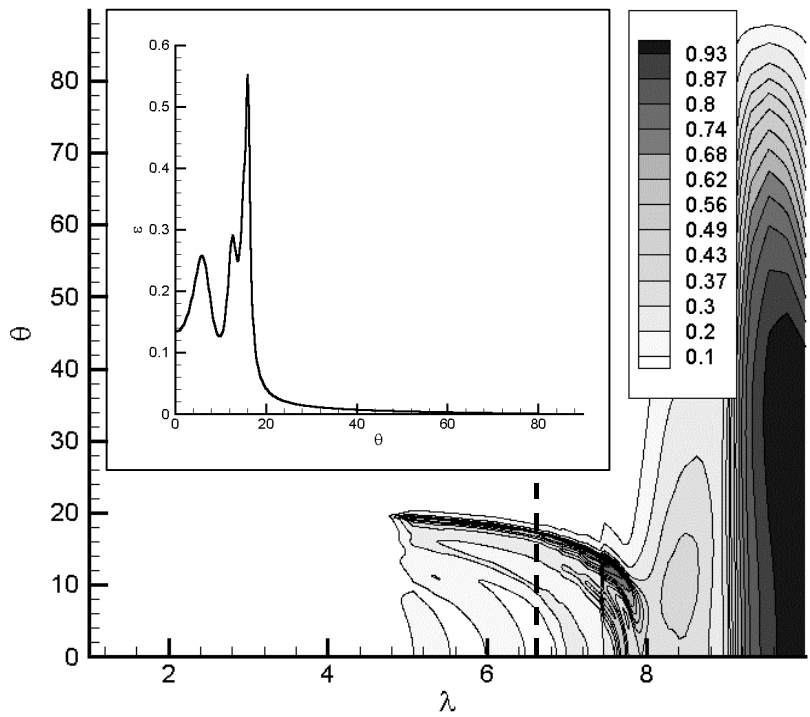

**Figure 2**

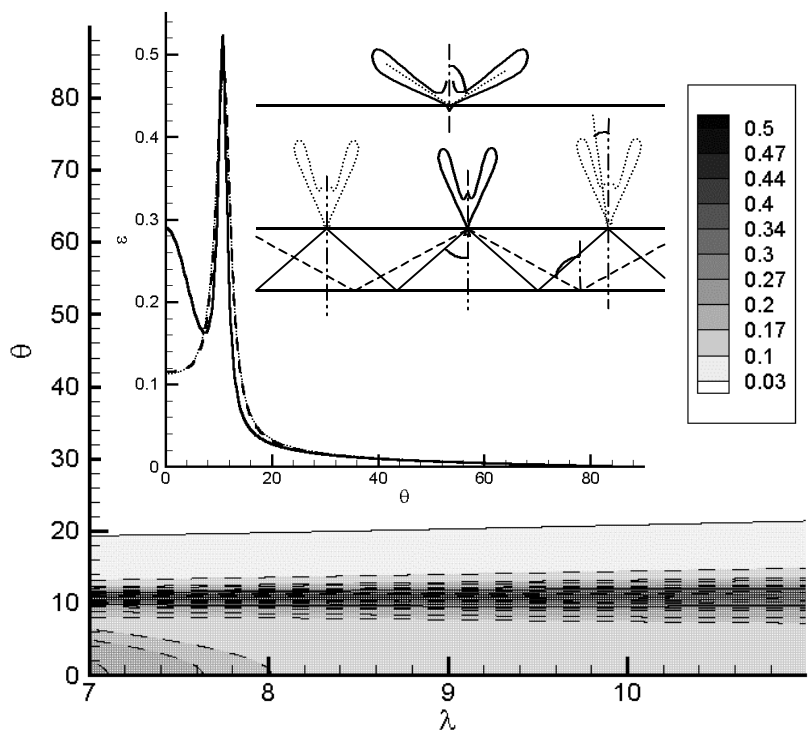

**Figure 3**